\documentclass[12pt,a4,sans]{article}
\usepackage[latin1]{inputenc}
\usepackage{amssymb}
\usepackage{amsmath}
\usepackage{amsthm}
\usepackage{amsfonts}
\usepackage[pdftex]{graphicx}
\usepackage{geometry}
\usepackage{bbold}
\usepackage{braket}
\usepackage{enumerate}
\usepackage[usenames,dvipsnames]{xcolor}
\usepackage{setspace}
\usepackage[backgroundcolor=white,linecolor=black]{todonotes}
\usepackage[in]{fullpage}
\usepackage{hyperref}
\usepackage{array}
\usepackage{cancel}
\usepackage{wrapfig}
\usepackage[font=small,labelfont=bf]{caption}
\usepackage{anysize}

\numberwithin{equation}{section}

\marginsize{2.4cm}{2.4cm}{2cm}{2cm}

\hypersetup{
colorlinks=true,
linktoc=page,
citecolor=Blue,
linkcolor=Blue,
urlcolor=Blue} 

\makeatletter 
\renewcommand{\maketitle} 
 { \begingroup \begin{center} \large {\bf \@title}
 	\vskip 5pt \large \@author \\ \vskip 5pt \@date \end{center}
   \vskip 5pt \endgroup \setcounter{footnote}{0} }
\makeatother 

\newcommand{\comments}[1]{}

\newcommand{\Tr}{\text{Tr}}

\newcommand{\black}{\mathord{\parbox[c]{1em}{\includegraphics[width=0.115\textwidth]{black}}}}

\newcommand{\be}{\begin{equation}}
\newcommand{\ee}{\end{equation}}

\def\beqa{\begin{eqnarray}}
\def\eeqa{\end{eqnarray}}
\def\beq{\begin{equation}}
\def\eeq{\end{equation}}

\def\Tr{{\rm Tr}}
\def\one{\mbox{1 \kern-.59em {\rm l}}}

\def\cA{{\cal A}} \def\cB{{\cal B}} \def\cC{{\cal C}}

 \def\cN{{\cal N}} \def\cO{{\cal O}}

\def\D{\Delta}

\def\uno{\mbox{1 \kern-.59em {\rm l}}}

\def\lan{\langle}
\def\ran{\rangle}

\def\one{1\!\!1\,\,}

\def\bcomment#1{}

\usepackage{slashed}
\usepackage{caption}

\usepackage[nottoc,notlot,notlof]{tocbibind}
\usepackage[nosort]{cite}
\usepackage{pstricks}         
\usepackage{color}
\usepackage{bbm}
\usepackage{parskip}
\usepackage{mathtools}

\graphicspath{{./Images/}}

\long\def\symbolfootnote[#1]#2{\begingroup%
\def\thefootnote{\fnsymbol{footnote}}\footnote[#1]{#2}\endgroup}

\begin{document}

\begin{flushright}
QMUL-PH-14-27
\end{flushright}

\vspace{20pt}

\begin{center}

{\Large \bf Integrability and MHV diagrams   }\\
\vspace{0.3 cm}
{\Large \bf in $\cN=4$ supersymmetric Yang-Mills theory   }

\vspace{45pt}

{\mbox {\bf Andreas Brandhuber, Brenda Penante, Gabriele Travaglini and Donovan Young}}%
\symbolfootnote[4]{
{\tt  \{ \tt \!\!\!\!\!\! a.brandhuber,  b.penante, g.travaglini, d.young\}@qmul.ac.uk}
}

\vspace{0.5cm}

\begin{center}
{\small \em

Centre for Research in String Theory\\
School of Physics and Astronomy\\
Queen Mary University of London\\
Mile End Road, London E1 4NS, UK
}
\end{center}


\vspace{40pt}

{\bf Abstract}
\end{center}

\vspace{0.3cm} 

\noindent

\noindent
We apply MHV diagrams to the  derivation of  the one-loop dilatation operator  of $\cN=4$ super Yang-Mills  in the  $SO(6)$ sector.  We find that in this  approach  the calculation reduces to the evaluation of a single MHV diagram in dimensional regularisation. This  provides the first application of MHV diagrams to an off-shell quantity. We also discuss other applications of the method and future directions.  
 \black
 
\setcounter{page}{0}
\thispagestyle{empty}
\newpage



%


\section{Introduction}

The study of $\mathcal{N}=4$  supersymmetric Yang-Mills (SYM) theory has led to the discovery of integrability in the planar limit,  providing the tools to compute the anomalous dimensions of local operators for any value of the coupling. In an initially independent line of research into this theory, the study of its on-shell scattering amplitudes has uncovered a rich structure and simplified calculations dramatically. It is  widely expected that the  integrability of the planar anomalous dimension problem and  the hidden structures and symmetries of scattering amplitudes  are related in some interesting way. In this paper we take a first step towards unravelling this potential connection.

Specifically, our goal here is to apply a method originally devised for computing amplitudes known as MHV diagrams \cite{Cachazo:2004kj} to  the derivation of the one-loop dilatation operator $\Gamma$ in the $SO(6)$ sector of $\cN=4$ SYM, originally computed by Minahan and Zarembo (MZ) in  \cite{Minahan:2002ve}. It is known that MHV diagrams are obtained from a particular axial gauge choice, followed by a field redefinition \cite{Mansfield:2005yd,Gorsky:2005sf}, hence the validity of the method not only applies to on-shell amplitudes, but also to off-shell quantities such as correlation functions. This paper  provides the first application of the MHV diagram method to the computation of correlation functions.

There are several reasons to pursue an approach based on MHV diagrams. Firstly, it is interesting to consider the application of this method to the computation of fully off-shell quantities such as correlation functions. Secondly, in the MHV diagram method  there is a natural way to regulate the divergences arising from loop integrations, namely dimensional regularisation, used  in conjunction with the  four-dimensional expressions for the vertices. In this respect, we recall that  one-loop amplitudes were calculated with MHV diagrams in  \cite{Brandhuber:2004yw}, where the infinite sequence of MHV amplitudes in $\cN=4$ SYM was rederived.  One-loop amplitudes in $\cN=1$ SYM were subsequently computed  in \cite{Bedford:2004py, Quigley:2004pw}, while in \cite{Bedford:2004nh} the cut-constructible part of the infinite sequence of MHV amplitudes in pure Yang-Mills  at one loop was presented. The $\cN=1$ and $\cN=0$ amplitudes have ultraviolet (UV) divergences (in addition to infrared ones), which are also regulated in dimensional regularisation. The two-point correlation function we  compute in this paper also exhibits  UV divergences, which we regulate  in exactly the same way as  in the case of amplitudes.%
\footnote{The reader may consult \cite{Brandhuber:2005kd,Brandhuber:2011ke} for further applications of the MHV diagram method to the calculation of loop amplitudes. }

An additional  motivation for our work is provided by the interesting recent papers \cite{Koster:2014fva,Wilhelm:2014qua}. In particular,  \cite{Koster:2014fva} successfully computed  $\Gamma$ using $\cN=4$ supersymmetric twistor actions  \cite{Boels:2006ir,Boels:2007qn,Adamo:2011cb}. It is known that such  actions, in conjunction with a particular axial gauge choice,  generate the MHV rules in twistor space \cite{Boels:2007qn}, and the question naturally arises as to whether one could derive the dilatation operator directly using MHV diagrams in momentum space, without passing through twistor space.  The answer to this question is positive, and furthermore we find that the calculation is very simple -- it amounts to the evaluation of a single MHV diagram in dimensional regularisation, leading to a single UV-divergent integral, identical to that appearing in  \cite{Minahan:2002ve}. 

The rest of the paper is organised as follows. In the next section we briefly review the result of \cite{Minahan:2002ve} for the integrable Hamiltonian describing the one-loop dilatation operator $\Gamma$  in the $SO(6)$ sector.  In Section 3 we address the calculation of $\Gamma$ using MHV diagrams. 
We present our conclusions and suggestions for future research in Section 4. 

\section{The one-loop dilatation operator} 

The computation of the dilatation operator in the $SO(6)$ sector of the $\cN=4$ SYM theory is equivalent to extracting the UV-divergent part of the two-point function $\left\langle \cO(x_1) 
\bar\cO(x_2)\right\rangle$, where $\cO$ is a single-trace scalar operator, of the form 
\beq
\cO_{A_1 B_1, A_2 B_2, \ldots, A_L B_L} (x) \ := \ \Tr \big( \phi_{A_1 B_1} (x) \cdots  \phi_{A_L B_L} (x) \big) \ . 
\eeq
At one loop and in the planar limit, only nearest neighbour scalar fields can be connected by vertices. This simplifies the calculation to that of 
$\left\langle (\phi_{AB} \phi_{CD})(x_1)(\phi_{A^\prime B^\prime} \phi_{C^\prime D^\prime})(x_2) \right\rangle $. 
The expected flavour structure of this correlation function is 
\begin{align}
\label{ABC}
\hspace{-0.3cm}
\left\langle (\phi_{AB} \phi_{CD})(x_1)(\phi_{A^\prime B^\prime} \phi_{C^\prime D^\prime})(x_2) \right\rangle \!=\!\cA\,\epsilon_{ABCD} \epsilon_{A^\prime B^\prime C^\prime D^\prime} + \cB\,  \epsilon_{AB A^\prime B^\prime} \epsilon_{CD  C^\prime D^\prime} + \cC\, \epsilon_{AB C^\prime D^\prime} \epsilon_{A^\prime B^\prime CD} . 
\nonumber\\
\end{align}
These three terms are usually referred to  as  trace, permutation and identity. We are only interested in computing the leading UV-divergent contributions to the coefficients $\cA$, $\cB$ and $\cC$, which according to \cite{Minahan:2002ve} are expected to be%
\footnote{In the definitions of  $\cA_{\rm UV}$, $\cB_{\rm UV}$, and $\cC_{\rm UV}$ we omit a factor of $\lambda/(8 \pi^2) \times \big(1 / ( 4 \pi^2 x_{12}^2)\big)^2 \times (1/ \epsilon)$.}
\beq
\cA_{\rm UV} \ = \ {1\over 2}\, , \qquad  \cB_{\rm UV} \ = \ -1 \, , \qquad \cC_{\rm UV} \ = \ 1 \ . 
\eeq
This leads to the famous result  of \cite{Minahan:2002ve} for the one-loop dilatation operator $\Gamma$  in the $SO(6)$ sector, 
\beq
\Gamma  \ = \  {\lambda \over 8 \pi^2} \sum_{n=1}^{L} \left( \uno \, - \, \mathbb{P}_{n, n+1} \, + \, {1\over 2} \, {\rm Tr}_{n, n+1}  \right) 
\ , 
\eeq
where $\mathbb{P}$ and ${\rm Tr}$ are the permutation and trace operators, respectively.  $L$ is the number of scalar fields in the operator, and $\lambda$ the 't Hooft coupling.

\begin{figure}[h]
\centering
\includegraphics[width=0.15\linewidth]{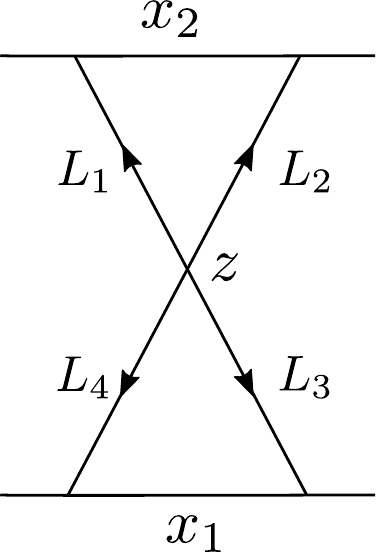}
\caption{\it The particular one-loop integral in configuration space contributing to the dilatation operator.}
\label{MZintegral}
\end{figure}

In the MZ calculation, one particular integral plays a central role, depicted in Figure \ref{MZintegral}. 
\\
It is given by
\begin{align}
\label{eq:X-position}
I(x_{12})\,=\,\int\!\!d^Dz \ \D^2 (x_1 - z) \, \D^2(x_2-z)\, , 
\end{align}
where $x_{12}:= x_1 - x_2$ and  
\beq
\D  (x)  \ := \ 
- {\pi^{2 - {D\over 2}} \over 4 \pi^2}
\Gamma \Big({D\over 2} - 1 \Big)
 {1\over (-x^2+ i \varepsilon)^{{D\over 2} - 1}}
\ , 
\eeq
is the scalar propagator in $D$ dimensions. 
Note that 
$I(x_{12})$  has UV divergences arising from the regions $z\rightarrow x_1$ and $z\rightarrow x_{2}$.

\begin{figure}[h]
\centering
\includegraphics[width=0.3\linewidth]{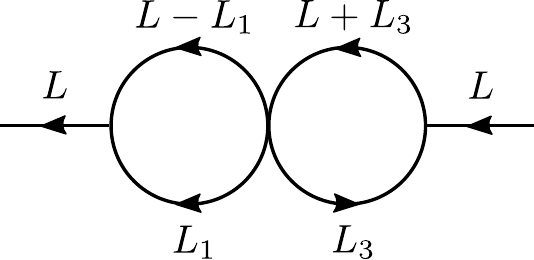}
\caption{\it The double-bubble integral relevant for the computation of $I(x_{12})$.}
\label{fig:double-bubble}
\end{figure}

Because the MHV diagram method is  formulated in momentum space, it is useful to recast $I(x_{12})$  as an integral in momentum space. Doing so one finds that 
\begin{align}
\begin{split}
I(x_{12})&= \int\!\prod_{i=1}^4 \frac{d^D L_i}{(2\pi)^D}\frac{e^{i(L_1+L_2)\cdot x_{12}}}{L_1^2\,L_2^2\,L_3^2\,L_4^2}\, (2 \pi)^D \, \delta^{(D)}\Big(\sum_{i=1}^4L_i\Big)
\\
& = \,\int\!\!\frac{d^D L}{(2\pi)^D} e^{iL\cdot x_{12}} \int\!\!\frac{d^D L_1}{(2\pi)^D}\frac{d^D L_3}{(2\pi)^D}\frac{1}{L_1^2\,(L-L_1)^2\,L_3^2\,(L+L_3)^2}\ , 
\label{eq:double-bubble}
\end{split}
\end{align}
where $L:= L_1 + L_2$. 
The integral over $L_1$ and $L_3$  is the product of two bubble integrals with momenta  as in Figure \ref{fig:double-bubble}, which are separately UV divergent. 
\\
These divergences arise  from the region $L_1$, $L_3\rightarrow \infty$. The leading UV divergence of \eqref{eq:double-bubble} is equal to 
\beq
\left. I(x_{12}) \right|_{\rm UV} \ = \ {1\over \epsilon}\cdot  {1\over 8 \pi^2} \cdot {1\over (4 \pi^2 x_{12}^2)^2} \ . 
\eeq

\section{The one-loop dilatation operator from MHV rules}

\begin{figure}[ht]
\begin{center}
\scalebox{0.33}{\includegraphics{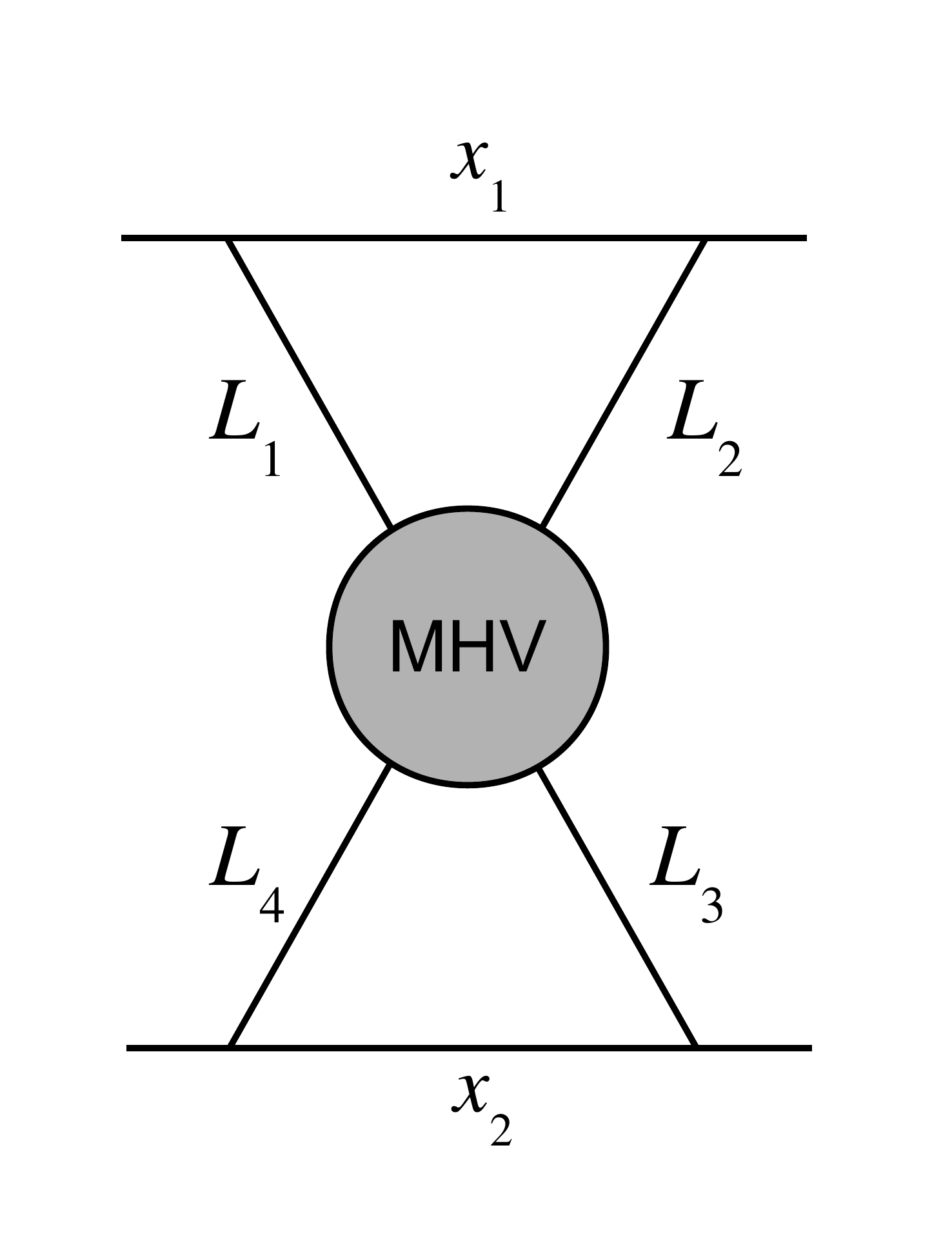}}
\end{center}
\caption{\it
The single MHV diagram contributing to the dilatation operator at one loop. 
}
\label{MHVdiagram}
\end{figure}

In this section we  compute the UV-divergent part of the coefficients $\cA$, $\cB$, $\cC$ defined in \eqref{ABC}, representing the trace, permutation and identity flavour structures, respectively.   
In order to compute these three coefficients, it is sufficient to consider one representative  configuration for each 
one.  We will choose the following helicity (or $SU(4)$) assignments: 
\begin{align}
\label{table1}
\begin{array}{c|cc}
 & ABCD & A'B'C'D' \\
 \hline
 \text{Tr} &1234 & 2413 \\
 \mathbb{P} & 1213 & 3424\\
 \uno & 1213 & 2434
\end{array}
\end{align}
There is a single   MHV diagram to compute, represented in Figure  \ref{MHVdiagram}. 
It consists of one  supersymmetric four-point MHV vertex, 
\beq
V_{\rm MHV} (1,2,3,4) \ = \ 
{ \delta^{(4)} (\sum_{i=1}^4 L_i ) \delta^{(8)} (\sum_{i=1}^4 \ell_i \eta_i   ) \over \lan 12\ran \lan 23\ran \lan 34\ran \lan 41\ran} \,  , 
\eeq
and four scalar propagators  $1 / (L_1^2 \cdots L_4^2)$ connecting it to the four scalars in the operators. 
Here 
$L_i$ are the (off-shell) momenta of the four particles in the vertex. The off-shell continuations of the spinors associated to the internal  legs are defined using the prescription of \cite{Cachazo:2004kj}, namely 
\beq
\ell_{i \alpha} :=   L_{i \alpha \dot{\alpha}} \xi^{\dot \alpha} 
\ . 
\eeq
Here $\xi^{\dot \alpha}$ is a constant reference spinor.%
\footnote{As we mentioned earlier, MHV diagrams were derived in \cite{Mansfield:2005yd,Gorsky:2005sf} from a change of variables in the Yang-Mills action quantised in the lightcone gauge. The  spinor  $\xi^{\dot \alpha}$ is precisely related to this gauge choice.}
Next we extract the relevant component vertices for the three flavour assignments in \eqref{table1}. These turn out to be: 
\beqa
{\rm Tr}:& \qquad \dfrac{ \lan 13\ran \lan 24\ran}{\lan 12 \ran \lan 34 \ran} \, , 
\nonumber \\ \cr
\mathbb{P}:& \qquad -1\, , 
\nonumber \\ \cr
\uno:& \qquad \dfrac{ \lan 13\ran \lan 24\ran}{\lan 23 \ran \lan 14 \ran} \, .
\label{sss}
\eeqa
Hence in the case of $\mathbb{P}$ the resulting loop integral is precisely the double-bubble  integral $I(x_{12})$ of  \eqref{eq:double-bubble} (up to a sign), while in the other two cases the double-bubble integrand is dressed with the vertex factors in \eqref{sss}. In the following we discuss the additional contributions from the vertex  for the three 
configurations ${\rm Tr}$, $\mathbb{P}$ and $\uno$. 

\subsection*{The Tr integrand} 

We begin our analysis with the vertex factor  for  the trace configuration, first line of \eqref{sss}.
Using the off-shell prescription for MHV diagrams we can rewrite  it as 
\beq
T:= { [\xi | L_1 L_3 |\xi]  \,  [\xi | L_2 L_4 |\xi] \over [\xi | L_1 L_2 |\xi] \,  [\xi | L_3 L_4 |\xi] }
\ . 
\eeq
Using momentum conservation to eliminate $L_2$ and $L_4$,  this can be recast as a sum of three terms, 
\beq 
\label{T2}
T \ = \  - { [\xi | L_1 L_3 |\xi]  \over   [\xi | L_3 L |\xi] } - { [\xi | L_1 L_3 |\xi]  \over   [\xi | L_1 L |\xi] }  - { [\xi | L_1 L_3 |\xi]^2   \over  [\xi | L_1 L |\xi] \,  [\xi | L_3 L |\xi] } 
\ , 
\eeq
where $L:= L_1 + L_2$. The first two terms correspond to linear bubble integrals in $L_1$ and $L_3$, respectively. We will study separately the contribution arising from the last term. 
The linear  bubble integral is 
\beq
\int\!{d^D K \over (2 \pi)^D} {K^\mu \over K^2 (K \pm L)^2} \ = \ \mp {L^\mu\over 2} {\rm Bub}(L^2)  \, ,
\eeq
where 
\beq
{\rm Bub} (L^2) := \int\!{d^D K \over (2 \pi)^D} {1 \over K^2 (K + L)^2}
\ . 
\eeq
This is one of the two scalar bubbles comprising   the MZ integral of  Figure \ref{fig:double-bubble}. In the following we will then only quote the coefficient dressing the MZ integral. Doing so, the first term in \eqref{T2} becomes, after the reduction,  
\beq 
- {[ \xi | L L_3 | \xi] \over [ \xi | L_3 L | \xi ] }\cdot {1\over 2}  \ = \ {1\over 2} \, .
\eeq
Similarly, the second term in \eqref{T2} gives a result of $+ 1/2$. Next we move to the third term. To simplify its expression, we first notice that the bubble integral in $L_1$ is symmetric under the transformation $L_1 \to L - L_1$. The idea is then to simplify the integrand by using this symmetry. Thus, we rewrite the quantity  $[ \xi | L_1 L_3 | \xi ] $ in the numerator as 
$[ \xi | L_1 L_3 | \xi ] \ = \ [ \xi  | (L_1 - {1\over 2} L) L_3 | \xi ] + {1\over 2} [\xi | L L_3 | \xi] $. Doing so, we get 
\beq
 - { [\xi | L_1 L_3 |\xi]^2   \over  [\xi | L_1 L |\xi] \,  [\xi | L_3 L |\xi] } =
 - { [\xi |( L_1  - {L\over 2}) L_3 |\xi]^2   \over  [\xi | L_1 L |\xi] \,  [\xi | L_3 L |\xi] } + {1\over 4}   { [\xi | L L_3 |\xi]   \over  [\xi | L_1 L |\xi]  } +   { [\xi |(L_1 - {1\over 2}  L)  L_3 |\xi]   \over  [\xi | L_1 L |\xi]  }\ . 
 \eeq
We then notice that the first  and the second term are  antisymmetric under the transformation $L_1 \to L - L_1$ and hence vanish upon integration. 
The third term is a sum of two  linear bubbles in $L_3$, and the corresponding contributions  are quickly seen to be equal to $- 1/2$ and zero, respectively. 

Summarising, the trace integral gives a contribution of $1/2$ times the dimensionally regularised   MZ integral. Thus $\cA_{\rm UV} =  {1/ 2}$.

\subsection*{The $\mathbb{P}$ integrand}
In this case the vertex is simply  $-1$ and the corresponding result is  $-1$ times the MZ integral, or $\cB_{\rm UV} = -1$. 

\subsection*{The $\uno$ integrand}

The relevant vertex factor is written in the third line of \eqref{sss}. In this case we observe that 
\beq 
{\lan 13 \ran \lan 24 \ran \over \lan 23 \ran \lan 14 \ran } \ = \  1 + {\lan 12 \ran \lan 34 \ran \over \lan 23 \ran \lan 14 \ran } 
 \ . 
 \eeq
 The first term gives a contribution equal to the MZ integral, and we will now argue that the second term is UV finite, and hence does not contribute to the dilatation operator. Indeed, we can write 
 \beq
 \label{sopra}
  {\lan 12 \ran \lan 34 \ran \over \lan 23 \ran \lan 14 \ran }  \ =\  {[ \xi | L_1 L | \xi ] [ \xi | L_3 L | \xi ] \over [ \xi |(L -  L_1) L_3 | \xi ] [ \xi | L_1 (L + L_3) | \xi ] }
  \ . 
  \eeq
  The UV divergences we are after arise when  $L_1$ and $L_3$ are large. The integrand \eqref{sopra} provides one extra power of momentum per integration, which makes each of the two bubbles in the MZ integral finite.%
 \footnote{One may also notice that for large $L_1$ and $L_3$ the integrand becomes an odd function of these two variables, and thus the integral should be suppressed even further than  expected from power counting.}   
 Thus  $\cC_{\rm UV}=1$.  
 
We end this section with a  couple of comments.  

{\bf 1.}   
Since MHV diagrams are obtained from a particular axial gauge choice, combined with a field redefinition \cite{Mansfield:2005yd,Gorsky:2005sf},  it is  guaranteed that $\xi$-dependence drops out at the end of the calculation. In the present case one can see this directly as follows. Lorentz invariance ensures that the result of the $L_1$- and $L_3$-integrations can only  depend on  $L^2$, as the other Lorentz-invariant quantity $[\xi | L^2 | \xi]$ vanishes (note that $L\cdot \xi$ cannot appear as our integrands only depend on the anti-holomorphic spinor $\xi_{\dot\alpha}$).
 
 {\bf 2.} 
 We  point out that  in the MHV diagram formalism there are no self-energy corrections to the propagator, as  already observed in Section 6 of \cite{Brandhuber:2004yw}.  Presumably this is also the case for the self-energy  evaluated with the twistor action of \cite{Boels:2007qn} employed in \cite{Koster:2014fva} for the calculation of the one-loop dilatation operator. It is  interesting to note that the superfield renormalisation in the lightcone gauge formalism of \cite{Belitsky:2004sc} is finite in the $\cN=4$ theory. 
  
\section{Conclusions} 

We conclude with some suggestions for future investigations. 

Firstly,  it would be interesting to apply MHV diagrams  to the calculation of the dilatation operator in other sectors of $\cN=4$ SYM, also containing fermions and derivatives.  Applications to different Yang-Mills theories with less supersymmetry can also be considered, given  the validity of the MHV diagram method   beyond $\cN=4$ SYM.

An obvious goal  is the extension of our calculation to  higher loops. This has proved difficult for amplitudes, but addressing the calculation of just the UV-divergent part of the two-point correlation function may simplify this task  enormously. At one loop the complete dilatation operator is known \cite{Beisert:2003jj}, while  
 direct perturbative calculations at higher loops   -- without the assumption of integrability --  have been performed only up to two   \cite{Eden:2005bt,Belitsky:2005bu,Georgiou:2011xj}, three  
\cite{Beisert:2003ys, Eden:2005ta,Sieg:2010tz} and four loops \cite{Beisert:2007hz} in particular sectors. A simplified route to such a calculation would be greatly desirable, and would provide further verification of this crucial assumption. The expected structure remains that of \eqref{eq:double-bubble}, with the double-bubble integral replaced by  more complicated (but still single-scale) loop integrals.

It would also be very interesting if one could  apply  other on-shell methods such as generalised unitarity  \cite{Bern:1997sc,Britto:2004nc} to  the direct calculation of  two-point functions, and hence to the dilatation operator of $\cN=4$ SYM.

Finally, our result hints at a link between the Yangian symmetry of amplitudes in $\cN=4$ SYM \cite{Drummond:2009fd} and integrability of the dilatation operator  of the theory 
\cite{Minahan:2002ve,Beisert:2003jj,Beisert:2003tq,Bena:2003wd, Beisert:2003yb,Beisert:2004ry}. It would be interesting to explore this point further. 

We hope to be able to report on some of these ideas in the very near  future.

\section*{Acknowledgements}

We would like to thank Massimo Bianchi, Rodolfo Russo, Bogdan Stefanski and Matthias Wilhelm for  interesting discussions. 
This work was supported by the Science and Technology Facilities Council Consolidated Grant ST/L000415/1  {\it ``String theory, gauge theory \& duality". }

\bibliographystyle{utphys}
\bibliography{dilatation}
\end{document}